\documentclass[journal]{IEEEtran}
\usepackage{mathrsfs}
\usepackage{bbm}
\usepackage{amssymb}
\usepackage{bbding}
\usepackage{threeparttable}
\usepackage{colortbl}
\usepackage[mathcal]{euscript}

\usepackage{psfrag,calc,url,bm}

\usepackage{cite}

\usepackage{graphicx}

\usepackage{psfrag}

\usepackage{subfigure}
\usepackage{hyperref}
\usepackage{url}

\usepackage{stfloats}

\usepackage{amsmath}

\usepackage{float}

\usepackage{algorithmic}

\usepackage[ruled,linesnumbered,vlined]{algorithm2e}

\usepackage{color}

\usepackage{boxedminipage}

\usepackage{amsthm}

\usepackage{multirow}

\usepackage{setspace}

\usepackage{soul}
\usepackage{epstopdf}

\usepackage{svg}

\allowdisplaybreaks[3]

\begin{document}

\title{KANsformer for Scalable Beamforming}

\author{Xinke Xie, Yang Lu, \IEEEmembership{Member, IEEE}, Chong-Yung Chi, \IEEEmembership{Life Fellow, IEEE},\\Wei Chen, \IEEEmembership{Senior Member, IEEE}, Bo Ai,~\IEEEmembership{Fellow,~IEEE}, and Dusit Niyato,~\IEEEmembership{Fellow,~IEEE}
\thanks{Xinke Xie and Yang Lu are with the School of Computer and Technology, Beijing Jiaotong University, Beijing 100044, China (e-mail: 21261022@bjtu.edu.cn,yanglu@bjtu.edu.cn).}
\thanks{Chong-Yung Chi is with the Institute of Communications Engineering, Department of Electrical Engineering, National Tsing Hua University, Hsinchu 30013, Taiwan (e-mail:cychi@ee.nthu.edu.tw).}
\thanks{Wei Chen and Bo Ai are with the School of Electronics and Information Engineering, Beijing Jiaotong University, Beijing 100044, China (e-mail: weich@bjtu.edu.cn,boai@bjtu.edu.cn).}
\thanks{Dusit Niyato is with the College of Computing and Data Science, Nanyang Technological University, Singapore 639798 (e-mail: dniyato@ntu.edu.sg).}
}

\maketitle

\begin{abstract}
This paper proposes an unsupervised deep-learning (DL) approach by  integrating transformer and Kolmogorov–Arnold networks (KAN) termed KANsformer to realize scalable beamforming for mobile communication systems. Specifically, we consider a classic multi-input-single-output energy efficiency maximization problem subject to the total power budget. The proposed KANsformer first extracts hidden features via a multi-head self-attention mechanism and then reads out the desired beamforming design via KAN. Numerical results are provided to evaluate the KANsformer in terms of generalization performance, transfer learning and ablation experiment. Overall, the KANsformer outperforms existing benchmark DL approaches, and is adaptable to the change in the number of mobile users with real-time and near-optimal inference.
\end{abstract}


\section{Introduction}

Deep learning (DL) has revolutionized a wide range of application fields and achieved unprecedented success in tasks such as image recognition and natural language processing. Its ability to automatically extract high-level features from raw data enables deep neural networks to outperform traditional machine learning methods in complex problem domains. Recently, the DL-enabled designs for wireless networks have emerged as a hot research topic\cite{AI}. Some researchers attempted to apply multi-layer perceptrons (MLP) \cite{mlp}, convolutional neural networks (CNN) \cite{cnn} and graph neural networks (GNN) \cite{lugnn} to deal with  the power allocation and signal processing problems in wireless networks. Overall, the DL models can be trained to achieve close performance to traditional convex optimization (CVXopt)-based approaches but with a much faster inference speed. How to further improve the learning performance remains an open issue for DL-enabled wireless optimization. 


Typically, task-oriented DL requires dedicated models for wireless networks. One promising way is to follow the ``encoder-decoder" paradigm, where the encoder extracts features over the wireless networks while the decoder maps the extracted features to desired transmit design. Recent works have paid a great attention to the design of encoder. Particularly, the GNN shows good scalability and generalization performance by exploiting the graph topology of wireless networks \cite{gnnshen}. In \cite{ligat,dl2,dl3}, the GNN was adopted as the encoder to develop the solution approaches for energy efficiency (EE) maximization, sum-rate maximization and max-min rate, respectively, for multi-user multi-input-single-output (MISO) networks, all of which were scalable to the number of users. In \cite{icnet}, a GNN based model was trained via unsupervised learning to solve the outage-constrained EE maximization problem. The GNNs in \cite{ligat,dl2,dl3,icnet} all leveraged the multi-head attention mechanism also known as the graph attention networks (GAT) to enhance the feature extraction, especially for inter-user interference. Similarly, the transformer is also built upon the self-attention mechanism and is popular for its highly predictive performance \cite{Transformer}. In \cite{zhutrans}, a transformer and a weighted A$^*$ based algorithm were proposed to plan the unmanned aerial vehicle trajectory for age-of-information minimization, which outperformed traditional algorithms numerically. However, the above works on wireless networks all adopted MLP as decoder. Recently, Kolmogorov-Arnold network (KAN) has been proposed as a promising alternative to MLP with superior performance in terms of accuracy and interpretability \cite{KAN}.

\emph{To the best of our knowledge, the integration of transformer and KAN has not been applied to beamforming design.} In this paper, we formulate the classic EE maximization problem for MISO networks \cite{ee}. We then propose an approach integrating transformer and KAN termed KANsformer, which utilizes the multi-head self-attention mechanism to extract hidden features among interference links and KAN to map the extracted features to the desired beamforming design. The KANsformer is trained via unsupervised learning and a scale function guarantees feasible solution. Via parameter sharing, the KANsformer is scalable to the number of users. Numerical results indicate that the KANsformer outperforms existing DL models and approaches the CVXopt-based solution accuracy with millisecond-level inference time. The major performance gain is contributed by the KAN via ablation experiment. Besides, we validate the scalability of KANsformer, which can be further enhanced by transfer learning at the expense of little training cost.

The rest of this paper is organized as follows. Section II gives the system model and problem formulation. Section III presents the structure of KANsformer. Section IV provides numerical results. Finally, Section V concludes the paper with future research directions.

\section{System Model and Problem Formulation}

Consider a downlink MISO network, where one $N_{\rm T}$-antenna transmitter intends to serve $K$ single-antenna mobile users (MUs) over a common spectral band. We use $\mathcal{K} \triangleq \{1,2, ..., K\}$ to denote the index set of the MUs.


Denote the symbol for the $k$-th MU and the corresponding beamforming vector as $s_k$ and ${{{\bf{w}}_k}}\in \mathbb{C}^{N_{\rm T}}$, respectively. The received signal at the $k$-th MU is given by
\begin{flalign}
{{\rm{y}}_k} = {\bf{h}}_k^H{{\bf{w}}_k}{s_k} + \sum\nolimits_{i \ne k}^K {{\bf{h}}_k^H{{\bf{w}}_i}{s_i}}  + {n_k},
\end{flalign}
where ${{{\bf{h}}_k}}\in{\mathbb{C}^{{N_{\rm{T}}}}}$ denotes the channel state information (CSI) of the $k$-th transmitter-MU link, and $n_k\sim\mathcal{CN}( {0,{\sigma_k ^2}})$ denotes the additive white Gaussian noise (AWGN) at the $k$-th MU. Without loss of generality, it is assumed that ${{\mathbb E}}\{ {{{| {s_k } |}^2}} \} = 1$ ($\forall k\in\mathcal{K}$). Then, the achievable rate at the $k$-th MU is expressed as
\begin{flalign}
{R_k}\left( {\left\{ {{{\bf{w}}_i}} \right\}} \right) = {\log _2}\left( {1 + \frac{{{{\left| {{\bf{h}}_k^H{{\bf{w}}_k}} \right|}^2}}}{{\sum\nolimits_{i=1, i \ne k}^K {{{\left| {{\bf{h}}_k^H{{\bf{w}}_i}} \right|}^2}}  + {\sigma_k ^2}}}} \right),
\end{flalign}
where $\{{\bf w}_i\}$ denotes the set of all admissible beamforming vectors. The weighted EE for the considered system is expressed as
\begin{flalign}
{\rm EE}\left( {\left\{ {{{\bf{w}}_i}} \right\}} \right) = \frac{{\sum\nolimits_{k = 1}^K \alpha_{k}{{R_k}\left( {\left\{ {{{\bf{w}}_i}} \right\}} \right)} }}{ \sum\nolimits_{k = 1}^K {\left\| {{{\bf{w}}_k}} \right\|_2^2}  + {P_{\rm C}}},
\end{flalign}
where $\alpha_{k}$ is a preassigned weight for the $k$-th MU and ${P_{\rm C}}$ denotes the constant power consumption introduced by circuit modules.

Our goal is to maximize the EE of the considered network, which is mathematically formulated as an optimization problem:
\begin{flalign}\label{p1}
{\left\{ {{{\bf{w}}^{\star}_i}} \right\}} = &\arg \mathop {\max }\limits_{\left\{{{{\bf{w}}_i}\in{\mathbb{C}^{{N_{\rm{T}}}}}} \right\},\sum\nolimits_{i=1}^K  \left\| {{{\bf{w}}_i}} \right\|_2^2 \le {P_{\rm max}}}{\rm{EE}}\left( {\left\{ {{{\bf{w}}_i}} \right\}} \right), 
\end{flalign}
where ${P_{\rm max}}$ denotes the power budget of the transmitter. 

The problem \eqref{p1} can be efficiently solved by existing CVX techniques but without close-form solution. By treating the CVXopt-based algorithm as a ``black box", it maps CSI to beamforming vectors via iterative computations. Such a mapping can also be regarded as a ``function" represented by $\Pi(\cdot):{\mathbb C}^{N_{\rm T}\times K}\rightarrow{\mathbb C}^{N_{\rm T}\times K}$. 
Following the universal approximation theorem, we intend to utilize neural networks to solve the problem \eqref{p1}.


\section{Structure of KANsformer}


The main idea of the proposed KANsformer is to realize the mapping, i.e., $\Pi(\cdot)$, from $\{{\bf h}_i\}$ to $\{{\bf w}_i\}$ such that ${\rm EE}( {\{ {{{\bf{w}}_i}} \}} )$ is close to ${\rm EE}( {\{ {{{\bf{w}}^{\star}_i}} \}} )$. Particularly, we utilize unsupervised learning to train the KANsformer to alleviate the burden on collecting labelled training set. Denote  $\bm\theta$ as learnable parameters of the KANsformer,  the loss function to update the learnable parameters is given by 
\begin{flalign}
{{\cal L} \left({\bm\theta}\right) = - {{\rm EE}\left(\Pi\left(\left\{{\mathbf{h}_i}\right\}\left|{\bm\theta}\right.\right)  \right)}}.
\end{flalign}
Note that via off-line training based on historical statistics, the KANsformer can derive the solution instantaneously at a low computational complexity instead of complex iterative calculation by the mathematical optimization approaches.

\begin{figure*}[htbp]
    \centering
\includegraphics[width=0.99\linewidth]{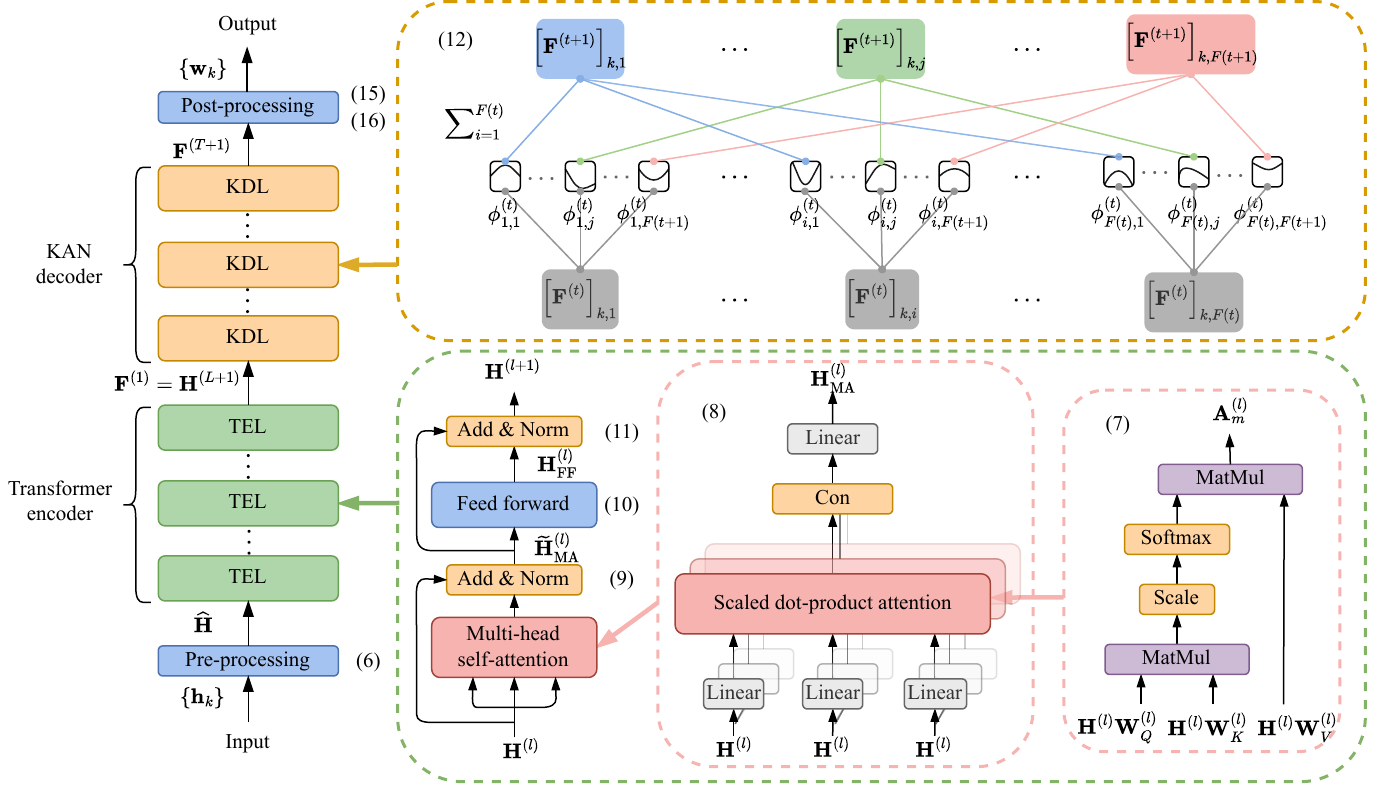}
    \caption{Structure of KANsformer which includes four modules: pre-processing module, transformer encoder module (with $L$ TELs), KAN decoder module (with $T$ KDLs) and post-processing module. The detailed processes of the $l$-th TEL and the $t$-th KDL are illustrated.}
    \label{KANsformer}
\end{figure*}

The structure of the KANsformer is illustrated in Fig.
 \ref{KANsformer}, which includes four modules: pre-processing module, transformer encoder module, KAN decoder module and post-processing module. The detailed processes in each module are described as follows.


\subsection{Pre-Processing Module}


In pre-processing module, we divide each complex-valued CSI vector in $\{{{\bf{h}}_k}\}$ into its real part, i.e., ${\rm Re}({{\bf{h}}_k})$, and its imaginary part, i.e., ${\rm Im}({{\bf{h}}_k}) $. Then, ${\rm Re}({{\bf{h}}_k})$ and ${\rm Im}({{\bf{h}}_k})$ are concatenated and input into a linear transformation to obtain the input for the transformer encoder module denoted by $\widehat{\bf H} \in \mathbb{R}^{K \times D}$, which given by
\begin{flalign}
    \widehat{\bf H}=&\left[{\rm{Con}}\left({\rm Re}\left({{\bf{h}}_1}\right),{\rm Im}\left({{\bf{h}}_1}\right)\right);\cdots;\right.\nonumber \\
    & \left.{\rm{Con}}\left({\rm Re}\left({{\bf{h}}_{K}}\right),{\rm Im}\left({{\bf{h}}_{K}}\right)\right)\right]{\bf W}_0,
\end{flalign}
where ${\rm Con}(\cdot)$ represents the concatenation operation, and ${\bf W}_0 \in \mathbb{R}^{2N_{\rm T}\times D}$ denotes the learnable parameters with $D$ being a configurable dimension which is usually greater than $K$ such that more attention heads can be employed.







\subsection{Transformer Encoder Module}

The aim of the transformer encoder module is to encode the obtained network feature $\widehat{\bf H}$ by exploring interactions among MUs and embedding the impact of inter-MU interference into the encoded network feature. The transformer encoder module comprises $L$ transformer encoder layers (TELs), each of which includes two ingredients, i.e., multi-head self-attention and position-wise feed-forward. For the $l$-th TEL, we denote its input and output{\footnote{The inputs and outputs of all the TELs have the same size.}} as ${\bf H}^{(l)}$ and ${\bf H}^{(l+1)} \in {\mathbb R}^{K \times D}$, respectively. The detailed processes of the two ingredients are given as follows. 



\subsubsection{Multi-Head Self-Attention}  Suppose that $M^{(l)}$ self-attention heads are employed in the $l$-th TEL, and then, the attention coefficient matrix associated with the $m$-th self-attention head in the $l$-th TEL is given by
\begin{flalign} 
    {\bf{A}}^{(l)}_m =&{\rm Softmax}\left(\frac{{\bf H}^{(l)}{\bf W}^{(l)}_Q\left({\bf H}^{(l)}{\bf W}^{(l)}_K\right)^T}{\sqrt{D}}\right){\bf H}^{(l)}{\bf W}^{(l)}_V\nonumber\\
    &\in {\mathbb R}^{K\times \frac{D}{M^{(l)}}},
\end{flalign}
where ${\bf W}^{(l)}_Q$, ${\bf W}^{(l)}_K$ and ${\bf W}^{(l)}_V \in \mathbb{R}^{D\times \frac{D}{M^{(l)}}}$ denotes the learnable parameters for the query, key and value projections, respectively.


The obtained $M^{(l)}$ attention heads, i.e., $\{{\bf A}_m^{(l)}\}_m$, are concatenated and then, passed into a linear layer with leanrable parameters of  ${\bf W}_{{\rm{MA}}}^{(l)} \in \mathbb{R}^{{D}\times {D}} $. Then, we obtain the multi-head attention coefficient matrix as
\begin{flalign}
    {\bf{H}}_{{\rm{MA}}}^{(l)} = \underbrace {{\rm{Con}}\left( {{\bf{A}}_{{1}}^{({{l}})} \cdots {\bf{A}}_{{{{M}}^{({{l}})}}}^{({{l}})}} \right)}_{\in\mathbb{R}^{K\times D}}{\bf{W}}_{{\rm{MA}}}^{({{l}})}\in \mathbb{R}^{K \times D}.
\end{flalign}  

To improve the training performance and stack deeper layers, the parameter-free layer normalization process represented by ${\rm LayerNorm}(\cdot)$ and the residual connection are adopted. The attention coefficient matrix is then updated by
\begin{flalign} 
    {\widetilde {\bf H}}_{\rm{MA}}^{(l)} = {\rm LayerNorm}\left({\bf H}_{{\rm{MA}}}^{(l)}\right) + {\bf H}^{(l)} \in \mathbb{R}^{K\times D}.
\end{flalign}

\subsubsection{ Position-wise Feed-forward Layer} The obtained attention coefficient matrix is input into a 2-layer feed-forward network with position-wise operation, i.e.,
\begin{flalign} 
    {\bf H}_{\rm{FF}}^{(l)} =&  {\rm Con}\left(f_2^{(l)}\left({\rm ReLu}\left( f_1^{(l)}\left( \left[{\widetilde {\bf H}}_{\rm{MA}}^{(l)}\right]_{k,:}\right)\right)\right)\right)\nonumber\\
    &\in \mathbb{R}^{K\times D},
    \end{flalign}
where $f_1^{(l)}(\cdot): \mathbb{R} ^{D}\rightarrow\mathbb{R} ^{ D^{\prime}}$ and $f_2^{(l)}(\cdot): \mathbb{R} ^{D^{\prime}}\rightarrow\mathbb{R} ^{ D}$ denote the feed-forward functions with $D^{\prime}$ being a intermediate dimension. 
The learnable parameters of $f_1^{(l)}\left(\cdot\right)$ and $f_2^{(l)}\left(\cdot\right)$ are denoted as ${{\bf W}^{(l)}_1}\in \mathbb{R}^{ D \times D^{\prime}} $ and ${{\bf W}^{(l)}_2}\in \mathbb{R}^{D^{\prime}\times D}$, respectively.

Similarly, the layer normalization process and the residual connection are followed by the feed-forward network, and the output of the $l$-th TEL is given by
\begin{flalign} 
    {\bf H}^{(l+1)} = {\rm LayerNorm}\left({\bf H}_{\rm{FF}}^{(l)}\right) + {{\bf H}_{\rm{FF}}^{(l)}}.
\end{flalign}


\subsection{KAN Decoder Module}

The aim of the KAN decoder module is to decode the obtained network features, i.e., ${\bf H}^{(L+1)}$, to the required beamforming vectors via $T$ KAN decoder layers (KDLs). For the $t$-th KDL, we denote its input and output as ${\bf F}^{(t)}\in {\mathbb R}^{K\times F(t)}$ and ${\bf F}^{(t+1)}\in {\mathbb R}^{K\times F(t+1)}$, respectively, where $F(t)$ and $F(t+1)$ denote the corresponding dimensions. Note that ${\bf F}^{(1)}={\bf H}^{\left(L+1\right)}$ and $F(1)=D$ while $F(T+1)=2N_{\rm T}$. The processing of the $t$-th KDL is given by
\begin{flalign} 
    &\left[ {\bf {\bf F}}^{(t+1)}\right]_{k,j}
    = \sum\nolimits_{i=1}^{F(t)} \phi_{j,i}^{(t)} \left(\left[ {\bf {\bf F}}^{(t)}\right]_{k,i}\right),
\end{flalign}
where $j\in \left\{1,..., F\left(t+1\right)\right\}$, $k\in\left\{1,...,K\right\}$ and $\phi_{j,i}^{(t)}(\cdot):{\mathbb R}\rightarrow{\mathbb R}$ is a continuous function which is given by
\begin{flalign} \phi_{j,i}^{(t)}\left(x\right)=\beta_{j,i}^{(t)} \frac{x}{1+\exp\left(-x\right)} + \gamma_{j,i}^{(t)} {\rm Spline}_{j,i}^{(t)}(x),
\end{flalign}
where $\beta_{j,i}^{(t)}$ and $\gamma_{j,i}^{(t)}$ are learnable parameters, and ${\rm Spline}_{j,i}^{(t)}(\cdot):{\mathbb R}\rightarrow {\mathbb R}$ is parameterized as a linear combination of B-splines such that
\begin{flalign}
    {\rm Spline}_{j,i}^{(t)}(x) = \sum\nolimits_{p=0}^{P} c_{p,j,i}^{(t)}B_p\left(x\right),
\end{flalign}
where $c_{p,j,i}^{(t)}$ denotes the learnable weights and  $P$ is a hyperparameter related to the B-splines (cf. \cite{B-spline}).

\subsection{Post-Processing Module}

The post-processing module is to convert ${{\bf F}^{(T+1)}}$ obtained by the KAN decoder module to the feasible solution to the problem \eqref{p1}. 

In particular, the real-valued ${\bf F}^{\left(T+1\right)}$ is used to recover $K$ complex-valued beamforming vectors with the $k$-th beamforming vector given by
\begin{flalign}
    \widetilde{{\bf w}}_k = {\bf F}^{\left(T+1\right)}\left[k,1:N_{\rm T}\right]+i{\bf F}^{\left(T+1\right)}\left[k,N_{\rm T}+1:2N_{\rm T}\right].
\end{flalign}

Then, each beamforming vector is fed into a scale function to satisfy the power budget of $P_{\rm max}$:
\begin{flalign}\label{af} 
 {{\bf{w}}}_k 
= \sqrt {\frac{{{P_{\rm max }}}} {{\rm max}\left({{P_{\rm max }}},\sum\nolimits_{i =1 }^{K} {{\left\| \widetilde{{\bf w}}_i \right\|}_2^2}\right)}}\widetilde{{\bf w}}_k. 
\end{flalign}

The learnable parameters in the KANsformer are given by 
\begin{flalign}
&\bm\theta = \left\{{{\bf W}_0},{\bf W}^{(l)},{\beta_{j,i}^{(t)}},{\gamma_{j,i}^{(t)}},{c_{p,j,i}^{(t)}}\right\},
\end{flalign}
where ${\bf W}^{(l)} \triangleq \{{{\bf W}^{(l)}_Q}, {{\bf W}^{(l)}_K},{{\bf W}^{(l)}_V},{{\bf W}^{(l)}_O},{{\bf W}^{(l)}_1},{{\bf W}^{(l)}_2}\} $.
Note that $\bm \theta$ is independent of $K$, thus facilitating the  KANsformer to accept the input $\{{{\bf{h}}_k}\}$ with different values of $K$.

\section{Numerical Results}
This section provides numerical results to evaluate the proposed KANsformer in terms of generalization performance, transfer learning and ablation experiment under the following settings.

\subsubsection{Simulation scenario} 
All the system parameters used are $N_{\rm T} \in \left\{4, 8, 16 \right\}$, $K \in \left\{ 2, 3, 4, 5, 7, 8, 9, 10, 12, 14\right\}$, $\alpha_k=1$ ($\forall k\in\mathcal{K}$), ${P_{\rm max}}=1$ W, ${P_{\rm C}}= 0.1$ W, CSI $\{ {\bf h}_{k}\in{\mathbb C}^{N_{\rm T}} \}$ being Rayleigh distributed for both training samples and test samples, and the corresponding labels (for test samples) representing the maximal EEs obtained by CVXopt-based algorithms. Specifically, we use $K_{\rm Tr}$, $K_{\rm Te}$ and $K_{\rm Tr}^\prime$ to respectively denote the number of MUs in the training stage, test stage and fine-tuning training stage (due to transfer learning), where $K_{\rm Te}\neq K_{\rm Tr}$ (known as scalability) is unknown during the training stage.

\subsubsection{Computer configuration} All DL models are trained and tested by Python 3.10 with Pytorch 2.4.0 on a computer with Intel(R) Xeon(R) Platinum 8255C CPU and NVIDIA RTX 2080 Ti (11 GB of memory).

\subsubsection{Initialization and training} The learnable parameters are initialized according to He (Kaiming) method and the learning rate is initialized as $10^{-4}$. The $Adam$ algorithm is adopted as the optimizer during the training phase. The batch size is set to $16$ for $100$ training epochs. The learnable weights with the best performance are used as the training results.

\subsubsection{Benchmark DL models}
In order to evaluate the KANsformer numerically, the following four baselines are considered, i.e.,
\begin{itemize}
\item {CVXopt-based approach}:  A single-layer successive convex approximation based optimization algorithm, similar to Algorithm 1 in \cite{lusee}, used to generate the test labels. 
\item {MLP}:  A basic feed-forward  neural network, similar to \cite{mlp}. 
\item {GAT}:  A basic GCN with multi-head attention mechanism, similar to \cite{ligat}.
\end{itemize}

\subsubsection{Test performance metrics} 
\begin{itemize}
  \item 
  Optimality performance: The ratio of the average achievable EE by the DL model to the optimal EE.
  \item 
  Inference time: Average running time for yielding the feasible beamforming solution by the DL model.
\end{itemize}

\begin{table}[t]
\centering
\caption{Generalization performance evaluation.}
\begin{tabular}{c|c|c||c|c|c|c}
\hline
$N_{\rm T}$ &$K_{\rm Tr}$&$K_{\rm Te}$ & CVX & MLP  & GAT & KF$^\dagger$\\
 \hline
 \hline
 \rowcolor{blue!10} 
\cellcolor{white}4&\cellcolor{white}2&\cellcolor{white}2&{100\%}&{98.2\%}&{ 98.4\%}&\bf{ 99.5\%}\\
 \hline
 \multicolumn{3}{c||}{Inference time}& 6.7s& \bf{3.3 ms}& 7.7 ms& 8.5 ms\\
 \hline
 \rowcolor{orange!10} 
\cellcolor{white}~&\cellcolor{white}~&\cellcolor{white}3&{$\times$}&{$\times$}&{ 84.3\%}& \bf{85.1\%}\\
 \rowcolor{blue!10} 
\cellcolor{white}8&\cellcolor{white}4&\cellcolor{white}4&{100\%}&{79.1\%}&{ 90.1\%}&\bf{95.3\%}\\
\rowcolor{orange!10} 
\cellcolor{white}~&\cellcolor{white}~&\cellcolor{white}5&{$\times$}&{$\times$}&{ 82.6\%}&\bf{83.2\%}\\
 \hline
 \multicolumn{3}{c||}{Inference time}& 10.5 s& \bf{3.4 ms}& 7.7 ms& 8.5 ms\\
 \hline
\rowcolor{orange!10}  
\cellcolor{white}~&\cellcolor{white}~&\cellcolor{white}7&{$\times$ }&{$\times$}&{84.0\%}&\bf{91.1\%}\\
 \rowcolor{blue!10} 
\cellcolor{white}16&\cellcolor{white}8&\cellcolor{white}8&{100\%}&{17.9\%}&{85.6\%}&\bf{92.9\%}\\
\rowcolor{orange!10} 
\cellcolor{white}~&\cellcolor{white}~&\cellcolor{white}9&{$\times$}&{$\times$}&{82.8\%}&\bf{90.8\%}\\
 \hline
 \multicolumn{3}{c||}{Inference time}& 57.4 s& \bf{3.5 ms}& 7.8 ms&8.4 ms\\
  \hline
\end{tabular}
\begin{tablenotes}
\footnotesize
\item $^\dagger$KF is short for KANsformer.
\item $\times$ represents ``not applicable".
\end{tablenotes}
\label{table:2}
\end{table}

\begin{table*}[ht]
\centering
\caption{Transfer learning evaluation: $N_{\rm T}=16$.}
\begin{tabular}{c||c|c|c|c| c}
\hline
 \multirow{2}*{$K_{\rm Te}$} & Scaling$^{\dagger}$ & Re-training ($K_{\rm Tr}=K_{\rm Te}$)  &  \multicolumn{3}{c}{Transfer learning ($K_{\rm Tr}=8$, $K'_{\rm Tr}=K_{\rm Te}$)}  \\
\cline{3-6}
 & ($K_{\rm Tr}=8$) & 100 epochs & 10 epochs & 20 epochs & 50 epochs\\
 \hline
 \hline
 10& \cellcolor{orange!10} {86.6\% }&\cellcolor{blue!10} {93.4\%}&{93.0\%} &\bf{93.5\%} & \bf{93.5\%}\\
 \hline
12&\cellcolor{orange!10} {77.9\%}&\cellcolor{blue!10} {90.7\%}&{94.6\%} & \bf{95.2\%} &\bf{95.2\%}\\
 \hline
14&\cellcolor{orange!10} {71.1\%}&\cellcolor{blue!10} {\bf{95.2\%}}&{93.6\%} &{94.2\%} &{94.2\%}\\
 \hline
\end{tabular}
  \begin{tablenotes}
\footnotesize
\item $^{\dagger}$Scaling represents that directly applying the model trained with $K_{\rm Tr}$ to the scenario of $K_{\rm Te}$.
\end{tablenotes}
\label{Transfer learning}
\end{table*}

\begin{table}[t]    
    \centering  
    \caption{Ablation experiment: $N_{\rm T}=16$ and $K_{\rm Tr}=8$.}
    \label{AblationPS}
    \begin{threeparttable}
    \begin{tabular}{cc|cc||c|c|c}  
 \hline
 \multicolumn{2}{c|}{Encoder} & \multicolumn{2}{c||}{Decoder}  &\multicolumn{3}{c}{$K_{\rm Te}$} \\
  \hline
      {GAT} & {TF}$^{\dagger}$ & {MLP} & {KAN} & $7$ & $8$ & $9$ \\
        \cline{5-7}
        \hline
        \hline
      $\checkmark$ & $\times$ & $\checkmark$ & $\times$  &\cellcolor{orange!10} 84.0\% &\cellcolor{blue!10} 85.6\% &\cellcolor{orange!10} 82.8\%  \\
        \cline{5-7}
       $\checkmark$ & $\times$ & $\times$  & $\checkmark$ &\cellcolor{orange!10} 89.5\% &\cellcolor{blue!10} 91.2\% &\cellcolor{orange!10} 88.1\%  \\
        \cline{5-7}
       $\times$ & $\checkmark$ &  $\checkmark$ & $\times$ &\cellcolor{orange!10} 82.5\% &\cellcolor{blue!10} 85.8\% & \cellcolor{orange!10} 83.2\%  \\
        \cline{5-7}
       $\times$ & $\checkmark$ & $\times$  & $\checkmark$ &\cellcolor{orange!10} {\bf 91.1\%} &\cellcolor{blue!10} {\bf 92.9\%} &\cellcolor{orange!10} {\bf 90.8\%}  \\
        \hline 
        \multicolumn{2}{c|}{Avg. gain} & \multicolumn{2}{c||}{-}&0.1\%&1.9\%& 3.1\%\\
        \hline
       \multicolumn{2}{c|}{-} & \multicolumn{2}{c||}{Avg. gain}&14.1\%&7.3\%&12.9\%\\
        \hline
    \end{tabular}
    \begin{tablenotes}
\footnotesize
\item $^{\dagger}$TF is short for transformer.
\end{tablenotes}
    \end{threeparttable}  
\end{table}  

\subsection{Generalization Performance}


The numerical performance tests and inference times of the KANsformer are given in Table \ref{table:2}, which are presented in more detail below, respectively.

\subsubsection{Optimality performance with $K_{\rm Te}=K_{\rm Tr}$ (marked by blue-shaded areas)} 

One can observe that the KANsformer outperforms the MLP and the GAT for all the three cases; the larger of $N_{\rm T}$ and $K_{\rm Te}=K_{\rm Tr}$, the larger the performance degradation, showing larger negative impact on the learning performance for all the DL models.  However, the KANsformer with the best performance maintains the performance loss within $10\%$.


\subsubsection{Optimality performance with $K_{\rm Te}\neq K_{\rm Tr}$  (marked by orange-shaded areas)} 


The KANsformer performs much better than the  GAT for $K_{\rm Tr}=8, /K_{\rm Te}\in\{7,9\}$, but slightly better for $K_{\rm Tr}=4,/K_{\rm Te}\in\{3,5\}$, besides some performance loss compared with the case for $K_{\rm Te}=K_{\rm Tr}\in\{4,8\}$. These results also indicate that the scalability performance loss is larger for larger $|K_{\rm Te}-K_{\rm Tr}|/K_{\rm Tr}$, because the multi-head self-attention mechanism intends to explore the
interaction among MUs, which may change with the number of MUs.

\subsubsection{Inference time} All of the MLP, GAT and KANsformer achieve millisecond-level inference (significantly faster than the iterative CVXopt-based approach) such that they are applicable under time-varying channel conditions. A more surprising observation is that the
inference time of DL models remains almost unchanged for all the numbers of $N_{\rm T}$ and $K$ used, while it increases exponentially for the CVXopt-based approach (a widely known fact). 


In summary, the well-trained KANsformer is able to achieve real-time and near-optimal inference for solving the problem \eqref{p1} while being scalable to the number of MUs (though $K_{\rm Te}$ unknown in the training stage) with an acceptable performance.


\subsection{Transfer Learning}

As mentioned that the scalability suffers from performance degradation with the increment of $|K_{\rm Te}-K_{\rm Tr}|/K_{\rm Tr}$. One can retrain the model or fine-tune the model via transfer learning on a new dataset (where the number of users is  $K_{\rm Te}$). The former initializes the learnable parameters randomly while the latter adopts the learnable parameters of the model trained for  $K_{\rm Tr}$ as the initial values of the model $\bm \theta$ instead.   Table \ref{Transfer learning} shows the performance of scaling, re-training (via $100$ epochs) and transfer learning (via $\{10,20,50\}$ epochs). It can be seen that the transfer learning can effectively improve the performance at a quite low training cost (e.g., $10$ epochs) compared with the performance of the plain scaling,  meanwhile achieving a comparable performance of the re-training at fewer training epochs (e.g., $20$ epochs). For $K_{\rm Te}=14$, the transfer learning falls behind the re-training by $1\%$, and the reason is that the prior-knowledge for $K_{\rm Tr}$ may mislead the transfer learning under $K^{\prime}_{\rm Te}=K_{\rm Tr}$ with large $|K^{\prime}_{\rm Tr}-K_{\rm Tr}|$. Nevertheless, the transfer learning can also achieve a considerable performance gain ($>20\%$) compared with the plain scaling.

\subsection{Ablation Experiment}

Table \ref{AblationPS} gives the ablation experiment to validate the effectiveness of the transformer used as the encoder and KAN used as the decoder. A performance gain can be  observed by comparing transformer/KAN and GAT/MLP for both cases of $K_{\rm Te} = K_{\rm Tr}$  and $K_{\rm Te}\neq K_{\rm Tr}$. Specifically, the average performance gains over $K_{\rm Te}\in\{7,8,9\}$ resulting from the transformer and  KAN are respectively $1.7\%$ and $11.4\%$. The reason for this is that both the GAT and the transformer adopt the attention mechanism to enhance the expressive capability while KAN has more flexible activation processes than MLP, such that KAN can outperform MLP in terms of interpretability \cite{KAN}. 

\section{Conclusion}



We have presented a DL model (i.e., the KANsformer shown in Fig. \ref{KANsformer}) with the transformer and KAN used in the encoder-decoder structure, respectively, for solving the beamforming design problem (cf. (\ref{p1})). 

Numerical results showed that the
KANsformer outperforms the existing DL models in terms of both the performance accuracy and the inference time consumed. Furthermore, we would like to emphasize that, in response to the given input CSI $\{{\bf h}_k\}$, the KANsformer can yield the beamforming vector output $\{{\bf w}_k\}$, with the elapsed inference time almost fixed (thus insensitive to the problem size) and tremendously lower than the problem-size dependent running time required by the CVXopt-based approach; the performance accuracy of the former is quite close to that the latter (treated as the optimum). These results also motivate further development of more powerful encoders and decoders dedicated to wireless communication systems.

\end{document}